\documentclass[aps,11pt]{revtex4}
\def\s{\sigma}

\def\ab{\overline{a}}
\def\tr{{\rm tr}}
\def\d{{\rm d}}
\def\E{{\rm E}}
\def\ro{{\hat\rho}}
\def\Ao{{\hat A}}
\def\Po{{\hat P}}
\def\Io{{\hat I}}
\def\Uo{{\hat U}}
\def\Ro{{\hat R}}
\def\Pit{{\mit\Pi}}
\def\Pio{{\hat\Pit}}
\def\Sit{{\mit\Sigma}}
\def\Sio{{\hat\Sit}}
\def\ket#1{|#1\rangle}
\def\bra#1{\langle#1|}
\def\brak#1#2{\langle#1|#2\rangle}
\begin{document}
\title{WEAK MEASUREMENTS IN QUANTUM MECHANICS}
\author{Lajos Di\'osi}
\affiliation{
Research Institute for Particle and Nuclear Physics\\
H-1525 Budapest 114, POB 49, Hungary}
\date{\today}

\begin{abstract}
The article recapitulates the concept of weak measurement in its broader sense encapsulating the trade 
between asymptotically weak measurement precision and asymptotically large measurement statistics. 
Essential applications in time-continuous measurement and in postselected measurement are presented 
both in classical and in quantum contexts. We discuss the anomalous quantum weak value in postselected 
measurement. We concentrate on the general mathematical and physical aspects of weak measurements and 
we do not expand on their interpretation. Particular applications, even most familiar ones, are not 
subject of the article which was written for Elsevier's Encyclopedia of Mathematical Physics.

\end{abstract}
\maketitle

\section{Introduction}
In quantum theory, the mean value of a certain observable $\Ao$ in a (pure) quantum state $\ket{i}$ 
is defined by the quadratic form [\ref{Ai}]. 
\begin{equation}\label{Ai} 
\langle\Ao\rangle_i=:\bra{i}\Ao\ket{i}
\end{equation}
Here $\Ao$ is hermitian operator on the Hilbert space ${\cal H}$ of states. We use Dirac formalism. The above
mean is interpreted statistically. No other forms had been known to possess a statistical 
interpretation in standard quantum theory. One can, nonetheless, try to extend the notion of mean for 
normalized bilinear expressions (Aharonov, Albert and Vaidman, 1988).
\begin{equation}\label{Aw} 
A_{\rm w}=:\frac{\bra{f}\Ao\ket{i}}{\brak{f}{i}}
\end{equation}
However unusual is this structure, standard quantum theory provides a plausible statistical
interpretation for it, too. The two pure states $\ket{i},\ket{f}$ play the roles of the prepared initial 
and the postselected final states, respectively. The statistical interpretation relies
upon the concept of weak measurement. In a single weak measurement, the notorious decoherence is chosen 
asymptotically small. In physical terms: the coupling between the measured state and
the meter is assumed asymptotically weak. The novel mean value [\ref{Aw}] is called the (complex) weak value. 

The concept of quantum weak measurement (Aharonov, Albert and Vaidman, 1988) provides
particular conclusions on postselected ensembles. Weak measurements have been instrumental in
the interpretation of time-continuous quantum measurements on single states as well.  
Yet, weak measurement itself can properly be illuminated in the context of 
classical statistics. Classical weak measurement as well as postselection and time-continuous measurement 
are straightforward concepts leading to conclusions that are natural in classical statistics. 
In quantum context, the case is radically different and certain paradoxical conclusions follow from weak 
measurements. Therefore we first introduce the classical notion of weak measurement on postselected ensembles and, 
alternatively, in time-continuous measurement on a single state. Certain idioms from statistical physics 
will be borrowed and certain not genuinely quantum notions from quantum theory will be anticipated. 
The quantum counterpart of weak measurement, postselection, and continuous measurement will be presented
afterwards. The apparent redundancy of the parallel presentations is of reason: the reader
can separate what is common in classical and quantum weak measurements from what is genuine quantum.

\section{Classical weak measurement}
Given a normalized probability density $\rho(X)$ over the phase space $\{X\}$, which we
call the state, the mean value of a real function $A(X)$ is defined by eqn [\ref{Ar}].  
\begin{equation}\label{Ar}
\langle A\rangle_\rho=:\int \d X A\rho 
\end{equation}
Let the outcome of an (unbiased) measurement of $A$ be denoted by $a$. Its stochastic
expectation value $\E[a]$ coincides with the mean [\ref{Ar}].
\begin{equation}\label{Ea}
\E[a]=\langle A\rangle_\rho 
\end{equation} 
Performing a large number $N$ of independent measurements of $A$ on the elements of the
ensemble of identically prepared states, the arithmetic mean $\ab$ of the outcomes yields a reliable estimate 
of $\E[a]$ and, this way, of the theoretical mean $\langle A\rangle_\rho$. 

Suppose, for concreteness, the measurement outcome $a$ is subject to a gaussian stochastic error of standard 
dispersion $\s>0$. The probability distribution of $a$ and the update of the state corresponding to the
Bayesian inference are described by eqns [\ref{pa}] and [\ref{Bayes}], respectively.
\begin{eqnarray}\label{pa}
p(a)=\left\langle G_\s(a-A)\right\rangle_\rho\\ 
                \label{Bayes}
\rho\rightarrow\frac{1}{p(a)}G_\s(a-A)\rho
\end{eqnarray} 
Here $G_\s$ is the central gaussian distribution of variance $\s$.
Note that, as expected, eqn [\ref{pa}] implies eqn [\ref{Ea}]. 
Nonzero $\s$ means that the measurement is nonideal, yet the expectation value $\E[a]$ remains calculable 
reliably if the statistics $N$ is suitably large. 

Suppose the spread of $A$ in state $\rho$ is finite, according to eqn [\ref{D2A}].
\begin{equation}\label{D2A}
\Delta^2_\rho A=:\langle A^2\rangle_\rho-\langle A\rangle_\rho^2~~<~~\infty
\end{equation} 
Weak measurement will be defined in the asymptotic limit [\ref{weakN},\ref{D2}] where both the
stochastic error of the measurement and the measurement statistics go to infinity.
It is crucial that their rate is kept constant.
\begin{equation}\label{weakN}
\s,N~~\rightarrow~~\infty
\end{equation} 
\begin{equation}\label{D2}
\Delta^2=:\frac{\s^2}{N}={\rm const}
\end{equation} 
Obviously for asymptotically large $\s$, the precision of individual measurements becomes 
extremely weak. This incapacity is fully compensated by the asymptotically large
statistics $N$. In the weak measurement limit [\ref{weakN},\ref{D2}], the probability distribution 
$p_{\rm w}$ of the arithmetic mean $\ab$
of the $N$ independent outcomes converges to a gaussian distribution [\ref{pwab}]. 
\begin{equation}\label{pwab}
p_{\rm w}(\ab)\rightarrow G_\Delta\left(\ab-\langle A\rangle_\rho\right) 
\end{equation}
The gaussian is centered at the mean $\langle A\rangle_\rho$, and the variance of the gaussian 
is given by the constant rate [\ref{D2}]. Consequently, the mean [\ref{Ar}] is reliably calculable
on a statistics $N$ growing like $\sim\s^2$. 

With an eye on quantum theory, we
consider two situations of weak measurement in classical statistics: postselection and
time-continuous measurement.

\subsection{Postselection}
For the preselected state $\rho$, we introduce postselection via the real function $\Pit(X)$ 
where $0\leq\Pit\leq1$. 
The postselected mean value of a certain real function $A(X)$ is defined by eqn [\ref{PAr}].
\begin{equation}\label{PAr}
_\Pit\!\langle A\rangle_\rho=:\frac{\langle\Pit A\rangle_\rho}{\langle\Pit\rangle_\rho}
\end{equation}
The denominator $\langle\Pit\rangle_\rho$ is the rate of postselection.
Postselection means that after having obtained the outcome $a$ regarding the measurement of $A$, 
we measure the function $\Pit$, too, in ideal measurement with random outcome $\mit\pi$ upon which
we base the following random decision. With probability $\mit\pi$ we include the current $a$ into the 
statistics and we discard it with probability $1-\mit\pi$. Then the coincidence of $\E[a]$ and 
$_\Pit\!\langle A\rangle_\rho$, like in eqn [\ref{Ea}], remains valid. 
\begin{equation}\label{PEa}
\E[a]=_\Pit\!\langle A\rangle_\rho 
\end{equation}
Therefore a large ensemble of postselected states allows one to estimate the postselected mean 
$_\Pit\!\langle A\rangle_\rho$. 

Classical postselection allows introducing the effective postselected state 
by eqn [\ref{Pr}].
\begin{equation}\label{Pr}
\rho_\Pit=:\frac{\Pit\rho}{\langle\Pit\rangle_\rho}
\end{equation}
Then the postselected mean [\ref{PAr}] of $A$ in state $\rho$ can, by eqn [\ref{PArAPr}], be expressed 
as the common mean of $A$ in the effective postselected state $\rho_\Pit$.
\begin{equation}\label{PArAPr}
_\Pit\!\langle A\rangle_\rho=\langle A\rangle_{\rho_\Pit} 
\end{equation}

As we shall see later, quantum postselection is more subtle and can not be reduced to common
statistics, i.e., to that without postselection. The quantum counterpart of postselected mean
does not exist unless we combine postselection and weak measurement.

\subsection{Time-continuous measurement}
For time-continuous measurement, one abandons the ensemble of identical states. One supposes that
a single time-dependent state $\rho_t$ is undergoing 
an infinite sequence of measurements [\ref{pa},\ref{Bayes}] of $A$ 
employed at times $t=\delta t,t=2\delta t,t=3\delta t,\dots$. The rate $\nu=:1/\delta t$ goes to 
infinity together with the mean squared error $\s^2$. Their rate is kept constant.
\begin{equation}\label{weaknu}
\s,\nu~~\rightarrow~~\infty
\end{equation} 
\begin{equation}\label{g2}
g^2=:\frac{\s^2}{\nu}={\rm const}
\end{equation} 
In the weak measurement limit [\ref{weaknu},\ref{g2}], the infinite frequent weak measurements of $A$ 
constitute the model of time-continuous measurement. Even the weak measurements will significantly influence the 
original state $\rho_0$, due to the accumulated effect of the infinite many Bayesian updates [\ref{Bayes}].
The resulting theory of time-continuous measurement is described by coupled Gaussian processes [\ref{da}] and [\ref{dr}]
for the primitive function $\alpha_t$ of the time-dependent measurement outcome and, respectively, for 
the time-dependent Bayesian conditional state $\rho_t$.
\begin{eqnarray}\label{da}
\d\alpha_t=\langle A\rangle_{\rho_t}\d t+g \d W_t\\
                \label{dr}
\d\rho_t=g^{-1}\left(A-\langle A\rangle_{\rho_t}\right)\rho_t \d W_t
\end{eqnarray}
Here $\d W_t$ is the It\^o differential of the Wiener process.

Eqns [\ref{da},\ref{dr}] are the special case of the Kushner-Stratonovich equations of time-continuous Bayesian 
inference conditioned on the continuous measurement of $A$ yielding the time-dependent outcome value $a_t$.
Formal time-derivatives of both sides of eqn [\ref{da}] yield the heuristic eqn [\ref{at}].
\begin{equation}\label{at}
a_t=\langle A\rangle_{\rho_t}+g\xi_t
\end{equation}
Accordingly, the current measurement outcome is always equal to the current mean plus a term 
proportional to standard white noise $\xi_t$. This plausible feature
of the model survives in the quantum context as well. As for the other equation [\ref{dr}],
it describes the gradual concentration of the distribution $\rho_t$ in such a way that
the variance $\Delta_{\rho_t}A$ tends to zero while $\langle A\rangle_{\rho_t}$ tends to a random
asymptotic value. The details of the convergence depend on the character of the continously 
measured function $A(X)$. Suppose a stepwise $A(X)$ according to eqn [\ref{step}].
\begin{equation}\label{step}
A(X)=\sum_\lambda a^\lambda P^\lambda(X)
\end{equation}
The real values $a^\lambda$ are step heights all differing from each other.
The indicator functions $P^\lambda$ take values $0$ or $1$ and form a 
complete set of pairwise disjoint functions on the phase space.
\begin{eqnarray}\label{Ps}
&&\sum_\lambda P^\lambda\equiv1\\
&&P^\lambda P^\mu=\delta_{\lambda\mu}P^\lambda
\end{eqnarray}
In a single ideal measurement of $A$, the outcome $a$ is one of the $a^\lambda$'s singled out at random.
The probability distribution of the measurement outcome and the corresponding Bayesian update of the state
are given by eqns [\ref{p}] and [\ref{meas}], respectively.
\begin{eqnarray}\label{p}
p^\lambda&=&\langle P^\lambda\rangle_{\rho_0}\\
                \label{meas}
\rho_0&\rightarrow&\frac{1}{p^\lambda}P^\lambda\rho_0=:\rho^\lambda.
\end{eqnarray}
Eqns [\ref{da},\ref{dr}] of time-continuous measurement are a connatural time-continuous resolution of the 
`sudden' ideal measurement [\ref{p},\ref{meas}] in a sense that they reproduce it in the limit $t\rightarrow\infty$.
The states $\rho^\lambda$ are trivial stationary states of the eqn [\ref{dr}]. It can be shown that
they are indeed approached with probability $p^\lambda$ for $t\rightarrow\infty$.  

\section{Quantum weak measurement}
In quantum theory, states in a given complex Hilbert space $\cal H$ are represented by 
nonnegative density operators $\ro$, normalized by $\tr\ro=1$. Like the classical states $\rho$,
the quantum state $\ro$ is interpreted statistically, referring to an ensemble of states
with the same $\ro$. Given a hermitian operator $\Ao$, called observable,
its theoretical mean value in state $\ro$ is defined by eqn [\ref{QAr}].
\begin{equation}\label{QAr}
\langle\Ao\rangle_\ro=\tr(\Ao\ro) 
\end{equation}
Let the outcome of an (unbiased) quantum measurement of $\Ao$ be denoted by $a$. Its stochastic
expectation value $\E[a]$ coincides with the mean [\ref{QAr}].
\begin{equation}\label{QEa}
\E[a]=\langle\Ao\rangle_\ro 
\end{equation} 
Performing a large number $N$ of independent measurements of $\Ao$ on the elements of the
ensemble of identically prepared states, the arithmetic mean $\ab$ of the outcomes yields a reliable estimate 
of $\E[a]$ and, this way, of the theoretical mean $\langle\Ao\rangle_\ro$. If the measurement
outcome $a$ contains a gaussian stochastic error of standard dispersion $\s$ then the
probability distribution of $a$ and the update, called collapse in quantum theory, of the state are 
described by eqns [\ref{Qpa}] and [\ref{coll}], respectively.
(We adopt the notational convenience of physics literature to omit the unit operator $\Io$ from
trivial expressions like $a\Io$.)
\begin{eqnarray}\label{Qpa}
p(a)=\left\langle G_\s(a-\Ao)\right\rangle_\ro\\ 
                \label{coll}
\ro\rightarrow\frac{1}{p(a)}G_\s^{1/2}(a-\Ao)\ro G_\s^{1/2}(a-\Ao)
\end{eqnarray} 
Nonzero $\s$ means that the measurement is non-ideal but the expectation value $\E[a]$ 
remains calculable reliably if $N$ is suitably large. 

Weak quantum measurement, like its classical counterpart, requires finite spread of the observable
$\Ao$ on state $\ro$, according to eqn [\ref{QD2A}].
\begin{equation}\label{QD2A}
\Delta^2_\ro\Ao=:\langle\Ao^2\rangle_\ro-\langle\Ao\rangle_\ro^2~~<~~\infty
\end{equation} 
Weak quantum measurement, too, will be defined in the asymptotic limit [\ref{weakN}] introduced
for classical weak measurement. Single quantum measurements can no more distinguish between the 
eigenvalues of $\Ao$. Yet, the expectation value $\E[a]$ of the outcome $a$ remains calculable
on a statistics $N$ growing like $\sim\s^2$. 

Both in quantum theory and in classical statistics, the emergence of non-ideal measurements from
ideal ones is guaranteed by general theorems. For completeness of the present article, we prove the 
emergence of the non-ideal quantum measurement [\ref{Qpa},\ref{coll}] from the standard von Neumann 
theory of ideal quantum measurements (von Neumann, 1955).
The source of the statistical error of dispersion $\s$ is associated with the state $\ro_M$ in the
complex Hilbert space ${\cal L}^{2}$ of a hypothetic meter. Suppose $R\in(-\infty,\infty)$ is the position of the 
``pointer''. Let its
initial state $\ro_M$ be a pure central gaussian state of width $\s$, then the density
operator $\ro_M$ in Dirac position basis takes the form [\ref{rM}].
\begin{equation}\label{rM}
\ro_M=\int \d R\int \d R' G_\s^{1/2}(R)G_\s^{1/2}(R')\ket{R}\bra{R'}
\end{equation} 
We are looking for a certain dynamical interaction to transmit the ``value'' of the observable
$\Ao$ onto the pointer position $\Ro$. To model the interaction, we define the unitary
transformation [\ref{Uo}] to act on the tensor space ${\cal H}\otimes{\cal L}^{2}$.
\begin{equation}\label{Uo}
\Uo=\exp(i\Ao\otimes\hat K)
\end{equation} 
Here $\hat K$ is the canonical momentum operator conjugated to $\Ro$ by eqn [\ref{K}].
\begin{equation}\label{K}
\exp(ia\hat K)\ket{R}=\ket{R+a}
\end{equation} 
The unitary operator $\Uo$ transforms the initial uncorrelated quantum state 
into the desired correlated composite state [\ref{Sio}].
\begin{equation}\label{Sio}
\Sio=:\Uo\ro\otimes\ro_M\Uo^\dagger
\end{equation}
Eqns [\ref{rM}-\ref{Sio}] yield the expression [\ref{Sio1}] for the state $\Sio$. 
\begin{equation}\label{Sio1}
\Sio=\int \d R\int \d R' G_\s^{1/2}(R-\Ao)\ro G_\s^{1/2}(R'-\Ao)\otimes\ket{R}\bra{R'}\nonumber
\end{equation} 
Let us write the pointer's coordinate operator $\hat R$ into the standard form [\ref{Ro}] in Dirac position
basis.
\begin{equation}\label{Ro}
\hat R=\int \d a \ket{a}\bra{a}
\end{equation} 
The notation anticipates that, when pointer $\hat R$ is measured ideally, the outcome $a$ plays the role
of the non-ideally measured value of the observable $\Ao$. Indeed, let us consider the ideal von Neumann 
measurement of the pointer position on the correlated composite state $\Sio$. The probability of the outcome
$a$ and the collapse of the composite state are given by the standard eqns [\ref{Qpa1}] and 
[\ref{coll1}], respectively.
\begin{eqnarray}\label{Qpa1}
p(a)&=&\tr\left[(\Io\otimes\ket{a}\bra{a})\Sio\right]\\
                \label{coll1}
\Sio&\rightarrow&
\frac{1}{p(a)}\left[(\Io\otimes\ket{a}\bra{a})\Sio(\Io\otimes\ket{a}\bra{a})\right]
\end{eqnarray} 
We insert eqn [\ref{Sio1}] into eqns [\ref{Qpa1},\ref{coll1}]. 
Furthermore, we take the trace over ${\cal L}^{2}$ of both sides of eqn [\ref{coll1}]. 
In such a way, as expected, eqns [\ref{Qpa1},\ref{coll1}] of ideal measurement of $\Ro$ yield the 
earlier postulated eqns [\ref{Qpa},\ref{coll}] of non-ideal measurement of $\Ao$.

\subsection{Quantum postselection}
A quantum postselection is defined by a hermitian operator satisfying 
$\hat 0\leq\Pio\leq\Io$. The corresponding postselected mean value of a certain observable $\Ao$ 
is defined by eqn [\ref{QPAr}].
\begin{equation}\label{QPAr}
_\Pio\!\langle\Ao\rangle_\ro=:{\rm Re}\frac{\langle\Pio\Ao\rangle_\ro}{\langle\Pio\rangle_\ro}
\end{equation}
The denominator $\langle\Pio\rangle_\ro$ is the rate of quantum postselection.
Quantum postselection means that after the measurement of $\Ao$, we measure the observable $\Pio$ 
in ideal quantum measurement and we make a statistical decision on the basis of the outcome $\mit\pi$. 
With probability $\mit\pi$ we include the case in question into the statistics while we discard it
with probability $1-\mit\pi$.  By analogy with the classical case [\ref{PEa}], one may ask whether
the stochastic expectation value $\E[a]$ of the postselected measurement outcome does 
coincide with the mean [\ref{QPEa}]. 
\begin{equation}\label{QPEa}
\E[a]\stackrel{?}{=}_\Pio\!\langle\Ao\rangle_\ro 
\end{equation}
Contrary to the classical case, the quantum eqn [\ref{QPEa}] does not hold. The quantum counterparts 
of classical eqns [\ref{PEa}-\ref{PArAPr}] do not exist at all. Nonetheless, the quantum postselected mean 
$_\Pio\!\langle\Ao\rangle_\ro$ possesses statistical interpretation although restricted to the context 
of weak quantum measurements. In the weak measurement limit [\ref{weakN},\ref{D2}], 
a postselected analogue of classical eqn [\ref{pwab}] holds  for the arithmetic mean $\ab$ of 
postselected weak quantum measurements.
\begin{equation}\label{Qpwab}
p_{\rm w}(\ab)\rightarrow G_\Delta\left(\ab-_\Pio\!\langle\Ao\rangle_\ro\right) 
\end{equation}
The gaussian is centered at the postselected mean $_\Pio\!\langle\Ao\rangle_\ro$, and the variance of the 
gaussian is given by the constant rate [\ref{D2}]. Consequently, the mean [\ref{QPAr}] becomes calculable
on a statistics $N$ growing like $\sim\s^2$. 

Since the statistical interpretation of the postselected quantum mean [\ref{QPAr}] is only possible for
weak measurements therefore $_\Pio\!\langle\Ao\rangle_\ro$ is called the (real) weak value of $\Ao$.
Consider the special case when both the state $\ro=\ket{i}\bra{i}$ and the postselected operator 
$\Pio=\ket{f}\bra{f}$ are pure states.
Then the weak value $_\Pio\!\langle\Ao\rangle_\ro$ takes, in usual notations, a particular form [\ref{VQPAr}]
yielding the real part of the complex weak value $A_{\rm w}$ [\ref{Ai}].
\begin{equation}\label{VQPAr}
_f\!\langle\Ao\rangle_{i}=:{\rm Re}\frac{\bra{f}\Ao\ket{i}}{\brak{f}{i}}
\end{equation}
The interpretation of postselection itself reduces to a simple procedure. 
One performs the von Neumann ideal measurement of the hermitian projector $\ket{f}\bra{f}$, then one includes 
the case if the outcome is $1$ and discards it if the outcome is $0$. The rate of postselection is
$\vert\brak{f}{i}\vert^2$. We note that a certain statistical interpretation of ${\rm Im}A_{\rm w}$, too, 
exists although it relies upon the details of the `meter'. 

We outline a heuristic proof of the central eqn [\ref{Qpwab}].
One considers the non-ideal measurement [\ref{Qpa},\ref{coll}] of $\Ao$
followed by the ideal measurement of $\Pio$. 
Then the joint distribution of the corresponding outcomes is given by eqn [\ref{ppia}].
The probability distribution of the postselected outcomes $a$ is defined by eqn [\ref{pposta}], and takes the
concrete form [\ref{pposta1}]. The constant ${\cal N}$ assures normalization.
\begin{equation}\label{ppia}
p(\pi,a)=\tr\left( \delta(\pi-\Pio)G_\s^{1/2}(a-\Ao)\ro G_\s^{1/2}(a-\Ao) \right)
\end{equation}
\begin{equation}\label{pposta}
p(a)=:\frac{1}{{\cal N}}\int\pi p(\pi,a) \d\pi
\end{equation}
\begin{equation}\label{pposta1}
p(a)=:\frac{1}{{\cal N}}\left\langle G_\s^{1/2}(a-\Ao)\Pio G_\s^{1/2}(a-\Ao) \right\rangle_\ro
\end{equation}
Suppose, for simplicity, that $\Ao$ is bounded.
When $\s\rightarrow\infty$, eqn [\ref{pposta1}] yields the first two moments of the outcome $a$ as described
by eqns [\ref{Ea1}] and [\ref{Ea2}], respectively.  
\begin{eqnarray}\label{Ea1}
\E[a]&\rightarrow& _\Pio\langle\Ao\rangle_\ro\\
                \label{Ea2}
\E[a^2]&\sim&\s^2
\end{eqnarray}
Hence, by virtue of the central limit theorem, the probability distribution [\ref{Qpwab}] follows for the
average $\ab$ of postselected outcomes in the weak measurement limit [\ref{weakN},\ref{D2}]. 

\subsection{Quantum weak value anomaly}
Unlike in classical postselection, effective postselected quantum states can not be introduced. 
We can ask whether the eqn [\ref{QPr}] defines a correct postselected quantum state.
\begin{equation}\label{QPr}
\ro_\Pio^? =:{\rm Herm}\frac{\Pio\ro}{\langle\Pio\rangle_\ro}
\end{equation}
This pseudo-state satisfies the quantum counterpart of the classical eqn [\ref{PArAPr}].
\begin{equation}\label{QPArAPr}
_\Pio\!\langle\Ao\rangle_\ro=\tr\left(\Ao\ro_\Pio^? \right) 
\end{equation}
In general, however, the operator $\ro_\Pio^?$ is not a density operator since it may be indefinite.
Therefore eqn [\ref{QPr}] does not define a quantum state.  Eqn [\ref{QPArAPr}] does not guarantee that
the quantum weak value $_\Pio\!\langle\Ao\rangle_\ro$ lies within the range of the eigenvalues of the
observable $\Ao$.  

Let us see a simple example for such anomalous weak values in the two-dimensional Hilbert space. 
Consider the pure initial state given by eqn [\ref{rPauli}] and the postselected pure
state given by eqn [\ref{PPauli}], where $\phi\in[0,\pi]$ is a certain angular parameter. 
\begin{eqnarray}\label{rPauli}
\ket{i}&=&\frac{1}{\sqrt{2}}\left[\begin{array}{l}e^{i\phi/2}\\e^{-i\phi/2}\end{array}\right]\\
                \label{PPauli}
\ket{f}&=&\frac{1}{\sqrt{2}}\left[\begin{array}{l}e^{-i\phi/2}\\e^{i\phi/2}\end{array}\right]
\end{eqnarray}
The probability of successful postselection is $\cos^2\phi$. If $\phi\neq\pi/2$ then the postselected
pseudo-state follows from eqn [\ref{QPr}]. 
\begin{equation}\label{QPrPauli}
\ro_\Pio^? =\frac{1}{2}\left[\begin{array}{cc}1&\cos^{-1}\phi\\\cos^{-1}\phi&1\end{array}\right]
\end{equation}
This matrix is indefinite unless $\phi=0$, its two eigenvalues are $1\pm\cos^{-1}\phi$.
The smaller the postselection rate $\cos^2\phi$ the larger is the violation of the positivity
of the pseudo-density operator. Let the weakly measured observable take the form [\ref{Aox}]. 
\begin{equation}\label{Aox}
\Ao=\left[\begin{array}{cc}0&1\\1&0\end{array}\right]
\end{equation}
Its eigenvalues are $\pm1$. We express its weak value from eqns [\ref{VQPAr},\ref{rPauli},\ref{PPauli}] or,
equivalently, from eqns [\ref{QPArAPr}] and [\ref{QPrPauli}].
\begin{equation}
_f\langle\Ao\rangle_i=\frac{1}{\cos\phi}
\end{equation}
This weak value of $\Ao$ lies outside the range of the eigenvalues of $\Ao$. 
The anomaly can be arbitrary large if the rate $\cos^2\phi$ of postselection decreases.  

Striking consequences follow from this anomaly if we turn to the statistical interpretation. 
For concreteness, suppose $\phi=2\pi/3$ so that $_f\langle\Ao\rangle_i=2$. On average, seventy-five 
percents of the statistics $N$ will be lost in postselection.
We learned from eqn [\ref{Qpwab}] that the arithmetic mean $\ab$ of the postselected
outcomes of independent weak measurements converges stochastically to the weak value
upto the gaussian fluctuation $\Delta$, as expressed symbolically by eqn [\ref{sb}]. 
\begin{equation}\label{sb}
\ab=2\pm\Delta
\end{equation}
Let us approximate the asymptotically large error $\s$ of our weak measurements by $\s=10$ which
is already well beyond the scale of the eigenvalues $\pm1$ of the observable $\Ao$. The gaussian
error $\Delta$ derives from eqn [\ref{D2}] after replacing $N$ by the size
of the postselected statistics which is approximately $N/4$.
\begin{equation}
\Delta^2=400/N
\end{equation} 
Accordingly, if $N=3600$ independent quantum measurements of precision $\s=10$
are performed regarding the observable $\Ao$ then the arithmetic mean $\ab$ of the 
$\sim900$ postselected outcomes $a$ will be $2\pm0.33$. This exceeds significantly the largest
eigenvalue of the measured observable $\Ao$. Quantum postselection appears to bias the
otherwise unbiassed non-ideal weak measurements. 

\subsection{Quantum time-continuous measurement} 
The mathematical construction of time-continuous quantum measurement is similar to the classical
one. We consider the weak measurement limit [\ref{weaknu},\ref{g2}] of an infinite sequence of
non-ideal quantum measurements of the observable $\Ao$ at $t=\delta t,2\delta t,\dots$, on the
time-dependent state $\ro_t$. The resulting theory of time-continuous quantum measurement is
incorporated in the coupled stochastic eqns [\ref{Qda},\ref{Qdr}] for the primitive function 
$\alpha_t$ of the time-dependent outcome and for the conditional time-dependent state $\ro_t$,
respectively (Di\'osi, 1988).  
\begin{eqnarray}\label{Qda}
\d\alpha_t=&~&\langle\Ao\rangle_{\ro_t}\d t+g\d W_t\\
                \label{Qdr}
\d\ro_t=&-&\frac{1}{8}g^{-2}[\Ao,[\Ao,\ro_t]]\d t +\nonumber\\
        &+&g^{-1}{\rm Herm}\left(\Ao-\langle\Ao\rangle_{\ro_t}\right)\ro_t \d W_t
\end{eqnarray}

The eqn [\ref{Qda}] and its classical counterpart [\ref{da}] are perfectly similar.
There is a remarkable difference between eqn [\ref{Qdr}] and its classical counterpart [\ref{dr}]. 
In the latter, the stochastic average of the state is constant: $\E[\d\rho_t]=0$, expressing the fact that
classical measurements do not alter the original ensemble if we `ignore' the outcomes of the
measurements. On the contrary, quantum measurements introduce irreversible changes to the original 
ensemble, a phenomenon called decoherence in the physics literature. Eqn [\ref{Qdr}] implies 
the closed linear first-order differential eqn [\ref{me}] for the stochastic average of the
quantum state $\ro_t$ under time-continuous measurement of the observable $\Ao$.
\begin{equation}\label{me}
\frac{\d\E[\ro_t]}{\d t}=-\frac{1}{8}g^{-2}\left[\Ao,\left[\Ao,\E[\ro_t]\right]\right]
\end{equation}
This is the basic irreversible equation to model the gradual loss of quantum coherence (decoherence) 
under time-continuous
measurement. In fact, the very equation models decoherence under the influence of a large class of
interactions like, e.g., with thermal reservoirs or complex environments. In two-dimensional Hilbert
space, for instance, we can consider the initial pure state $\bra{i}=:[\cos\phi,\sin\phi]$ and the 
time-continuous measurement of the diagonal observable [\ref{Aoz}] on it. 
The solution of eqn [\ref{me}] is given by eqn [\ref{rt}].
\begin{equation}\label{Aoz}
\Ao=\left[\begin{array}{cc}1&0\\0&-1\end{array}\right]
\end{equation}
\begin{equation}\label{rt}
\E[\ro_t]=
\left[\begin{array}{cc}\cos^2\phi&e^{-t/4g^2}\cos\phi\sin\phi \\e^{-t/4g^2}\cos\phi\sin\phi&\sin^2\phi\end{array}\right]
\end{equation}    
The off-diagonal elements of this density matrix go to zero, i.e., the coherent superposition represented
by the initial pure state becomes an incoherent mixture represented by the diagonal density matrix $\ro_\infty$.

Apart from the phenomenon of decoherence, the stochastic equations show remarkable similarity with the
classical equations of time-continuous measurement. The heuristic form of eqn [\ref{Qda}] is eqn [\ref{Qat}]
of invariable interpretation with respect to the classical eqn [\ref{at}]. 
\begin{equation}\label{Qat}
a_t=\langle\Ao\rangle_{\ro_t}+g\xi_t
\end{equation}
Eqn [\ref{Qdr}] describes what is called the time-continuous collapse of the quantum state under 
time-continuous quantum measurement of $\Ao$. 
For concreteness, we assume discrete spectrum for $\Ao$ and consider the spectral expansion [\ref{spect}]
\begin{equation}\label{spect}
\Ao=\sum_\lambda a^\lambda\Po^\lambda
\end{equation}
The real values $a^\lambda$ are nondegenerate eigenvalues.
The hermitian projectors $\Po^\lambda$ form a complete orthogonal set.
\begin{eqnarray}\label{QPs}
&&\sum_\lambda\Po^\lambda\equiv\Io\\
&&\Po^\lambda\Po^\mu=\delta_{\lambda\mu}\Po^\lambda
\end{eqnarray}
In a single ideal measurement of $\Ao$, the outcome $a$ is one of the $a^\lambda$'s singled out at random.
The probability distribution of the measurement outcome and the corresponding collapse of the state
are given by eqns [\ref{Qp}] and [\ref{Qmeas}], respectively.
\begin{eqnarray}\label{Qp}
p^\lambda&=&\langle\Po^\lambda\rangle_{\ro_0}\\
                \label{Qmeas}
\ro_0&\rightarrow&\frac{1}{p^\lambda}\Po^\lambda\ro_0\Po^\lambda=:\ro^\lambda
\end{eqnarray}
Eqns [\ref{Qda},\ref{Qdr}] of continuous measurements are an obvious time-continuous resolution of the `sudden' ideal
quantum measurement [\ref{Qp},\ref{Qmeas}] in a sense that they reproduce it in the limit $t\rightarrow\infty$.
The states $\ro^\lambda$ are stationary states of the eqn [\ref{Qdr}]. It can be shown that
they are indeed approached with probability $p^\lambda$ for $t\rightarrow\infty$ (Gisin, 1984).  

\section{Related contexts}
In addition to the two particular examples as in postselection and in time-continuous measurement, respectively
presented above, the weak measurement limit itself has further variants. A most natural example is
the usual thermodynamic limit in standard statistical physics. Then weak measurements concern a certain additive
microscopic observable (e.g.: the spin) of each constituent and the weak value represents the corresponding 
additive macroscopic parameter (e.g.: the magnetization) in the infinite volume limit. This example indicates
that weak values have natural interpretation despite the apparent artificial conditions of their definition.
It is important that the weak value, with or without postselection, plays the physical role similar to that 
of the common mean $\langle\Ao\rangle_\ro$. If, between their pre- and postselection, the states $\ro$ become 
weakly coupled with the state of another quantum system via the observable $\Ao$ their average influence will be 
as if $\Ao$ took the weak value $_\Pio\langle\Ao\rangle_\ro$.   
Weak measurements also open a specific loophole to circumvent quantum limitations 
related to the irreversible disturbances that quantum measurements cause to the measured state.    
Non-commuting observables become simultaneously measurable in the weak limit: simultaneous weak values
of non-commuting observables will exist. 

Literally, weak measurement had been coined in 1988 for quantum measurements with (pre- and) postselection, and
became the tool of a certain time-symmetric statistical interpretation of quantum states. Foundational applications 
target the paradoxical problem of pre- and retrodiction in quantum theory. In a broad sense, however, 
the very principle of weak measurement 
encapsulates the trade between asymptotically weak precision and asymptotically large statistics. Its relevance in 
different fields has not yet been fully explored and a growing number of foundational, theoretical, and experimental 
applications are being considered in the literature -- predominantly
in the context of quantum physics. 
Since specialized monographs or textbooks on quantum weak measurement are not yet available the reader is 
mostly referred to research articles, like the recent one by Aharonov and Botero (2005), covering many
topics of postselected quantum weak values.  

This work was prepared for Encyclopedia of Mathematical Physics (Elsevier). 
The author was also supported by the Hungarian OTKA Grant No. 49384.

\end{document}